# Effect of Background Signal on Momentum Imaging

Sukanta Das, Suvasis Swain, Krishnendu Gope, Vishvesh Tadsare and Vaibhav S. Prabhudesai

Tata Institute of Fundamental Research, Colaba, Mumbai 400005 India

*vaibhav@tifr.res.in

The velocity Slice Imaging technique has revolutionised electron molecule interaction studies. Multiple electrostatic lens assemblies are often used in spectrometers for resolving low kinetic energy fragments. However, in a crossed-beam experiment with an effusive molecular beam, the extended source of ion generation due to the presence of the background gas creates artefacts on the momentum images as we try to magnify them beyond a certain size. Here, we present a systematic study of this effect on momentum imaging and the solutions to address this issue by background subtraction with suitable magnification. Additionally, we demonstrated that a supersonic molecular beam target helps minimise these artefacts in the image magnification by reducing the background signal. These systematic findings may bring valuable insight into the investigation of low kinetic energy release processes involving electron impact, ion impact, and merge beam experiments with large interaction volumes where high magnification is needed.

## I. INTRODUCTION

In the molecular collision process, the details of the dynamics leading to the dissociation are carried away by the fragments generated in a single collision condition. By measuring the momenta of these fragments, one can identify different reaction paths leading to the process of one's interest and unravel the role of molecular dynamics in the final outcome. Usually, ion imaging techniques are used to capture these momentum distributions. Chandler and Houston first demonstrated ion imaging on a 2D detector[1]. Later it was improved by Eppink and Parker[2] to Velocity Map Imaging (VMI) which reduces the effect of spatial spread of the interaction region on the final momentum image. Offerhaus *et al.* introduced a three-lens system at the entrance of the drift tube to magnify the momentum image of the low-energy electrons and ions[3]. They have shown a 20x magnification of the slow photoelectron emitted from photoionisation of Xe metastable state. These magnified images opened paths of molecular microscopy[4-6], which was not possible before. But the VMI technique required reconstruction of a 3D image from its 2D projection, for which different methods like able inversion[7,8], Onion peeling[9], iterative inversion[10], BASEX[11], and pBASEX[12] were utilised. All these methods required a cylindrical symmetry about the axis parallel to the detector plane. These techniques are very prone to noise and often leave a noisy patch along the line of symmetry[7, 8, 10] or at the centre[9, 11] of the image. In recent times, Sparling *et al.* have developed a new method of image reconstruction using an artificial neural network[13, 14]. This method does not require the presence of cylindrical symmetry.

Kitsopoulos and co-workers developed another imaging method[15] called Velocity Slice Imaging (VSI). In this method, instead of detecting the entire Newton sphere for the ion cloud, only its central slice is detected. Here, no cylindrical symmetry is required to obtain the momentum distribution of the ions generated. This method does not require any inversion algorithms and provides cleaner images. In VMI, a very high extraction voltage is given to pancake the image on the 2D detector. On the other hand, in VSI, the ion cloud is stretched in time. Kitsopoulos *et al.* used delayed pulsed extraction and a wire mesh on the extractor to keep the region between the repeller and extractor field free during the expansion. Later, Suits *et al.*[16] implemented a new design similar to VMI with lower extraction voltage in DC mode. They also used an array of lenses to stretch the molecular cloud in time inside the acceleration region, improving the image resolution. VSI technique gained rapid popularity among crossed molecular beam experiments. Lin *et al.*[17] used it for the first time for crossed molecular beam experiment. Nandi *et al.* adapted this technique for low-energy electron collision experiments[18]. Later several groups[19-23] used this technique in dissociative electron attachment (DEA) experiments. Over time, different modifications are made to improve the resolution of direct and sliced imaging techniques[24-27].

Throughout the last three decades, these imaging techniques have been optimised for better resolution, minimising the noise, and increasing the magnification capability while maintaining the VMI condition to study very low-energy ions or electrons. In all these experiments, crossed-beam geometry is used for creating the interaction volume. Typically, for the photodissociation and photoionisation experiments, the light beam is focused in the interaction region, confining the interaction volume to its Rayleigh range. Unlike these experiments, the projectile beam is not necessarily focused in the charged particle interaction experiments. This beam interacts with the background gas along its path in addition to the relatively denser target beam. This results in a far-extended spatial spread in the interaction volume. In such cases, the ion momentum imaging spectrometer has to handle this extended spatial spread in the ion generation and keep reasonable imaging resolution using electrostatic lenses. For an effusive molecular beam, the density of the background gas that comprises mainly the target molecules is comparable to the in-beam target density. As a result, the non-negligible contribution from this background is difficult to eliminate from the measured momentum image. However, this extended region of ion generation affects the quality of the image and acts as a source of noise, degrading the imaging resolution. This effect becomes adverse when the momentum images are magnified beyond a certain size, especially for the processes with low kinetic energy release, where magnification is necessary to resolve the image.

Here we show that one can obtain the optimised imaging condition for a given initial momentum distribution where the extended volume of the ion generation does not limit the quality of the momentum image. However, these optimised conditions are very specific to the initial kinetic energy magnitude as well as the magnification of the image. On the other hand, by subtracting the background signal directly, one can suppress this effect to a reasonable extent. We also show that using a supersonic molecular beam target reduces this effect substantially for a much larger range of image magnification.

## II. EXPERIMENTAL SETUP

We have used two VSI spectrometers with two types of target-generating mechanisms. The first set-up uses the effusive molecular beam from a capillary array, while the second one uses the supersonic molecular expansion from a multistage skimmer assembly to prepare the molecular target. The details of these setups are given below.

### A. Set-up 1

Figure 1 (a) shows the VSI spectrometer with four lens electrodes, the details of which are given elsewhere[28]. Here, we briefly describe it. This setup consists of an interaction region spanned by a set of two electrodes, namely a pusher and a puller, separated by 20 mm. The puller electrode has a molybdenum wire mesh with 64% transmittance to prevent the field penetration of the accelerating potentials into the interaction region. The acceleration region consists of a four-elements-electrostatic lens assembly to guide the ions through a short flight tube (10 mm long) towards the detector. The first and third lens electrodes are 6 mm thick, and the second and fourth are 2 mm thick. The separation among various electrodes is shown in the figure. All the lens electrodes, along with the flight tube entrance, have a 40 mm diameter aperture. The lens assembly is used to control the size and space focusing of the ion cloud by adjusting the potential on each of the electrodes. The spectrometer uses a molecular target prepared using an effusive molecular beam generated by a capillary array of length 10 mm. Each capillary of the array has a 100 μm diameter. The molecular beam is coaxial with the spectrometer axis. We term this mode of operation the crossed-beam mode. A Granville Phillips gas regulator is used before the capillary to introduce the gas inside the chamber in a regulated manner for maintaining the effusive flow condition where intra-molecular collision is negligible compared to collision with the wall. An MKS Baratron pressure gauge is used to measure the pressure behind the capillary. The chamber pressure is measured using an ionisation gauge (Granville Phillipe). The setup also has an arrangement for filling the vacuum chamber with the target gas at a given pressure. We term it as a static gas mode operation. The low-energy electron beam is produced using a home-built thermionic electron gun. The gun operates in a pulsed mode with an adjustable repetition rate in the range 100Hz to 10kHz. The electron beam is collimated in the interaction region using a pair of magnet coils mounted in the Helmholtz geometry outside the vacuum chamber. The electron current is measured using the home-built Faraday cup, mounted coaxially with the electron gun on the opposite side of the interaction region, as shown in Figure 1 (a). A 2D position-sensitive microchannel plate (MCP) based detector in a chevron configuration is used to detect the ions. The MCP detector is followed by the Phosphor screen (P43). A CCD camera mounted outside the chamber is used to capture the images of the Phosphor screen. The position information of the ion hits is determined in the offline analysis of the captured images using programs written in Matlab.

Typical operating pressure is a few hundred of mTorr behind the capillary, resulting in the background pressure of 5 x $10^{-7}$ to 1 x $10^{-6}$ Torr in the VSI spectrometer region.

## B. Setup-2

The schematics of this setup are shown in Figure 1(b). The spectrometer used in this setup is similar to setup1, except that the interaction region has two additional ring electrodes mounted at 5mm from the pusher and puller electrodes each. The potential divider arrangement across the pusher and puller electrodes via these ring electrodes is used to apply a uniform electric field across the interaction region. The puller electrode is equipped with a molybdenum wire mesh of 64% transmittance. The lens electrodes have varying apertures of diameter 34 mm, 36 mm, 38 mm, and 40 mm, separated by 5 mm from each other with a thickness of 1 mm, followed by an 80 mm long Flight tube.

Here, a two-stage skimmer assembly is used to prepare the supersonic molecular target beam coaxial to the VSI spectrometer axis. The setup consists of three vacuum chambers. These chambers are separated by conical skimmers of diameter 1 mm with a separation of 100 mm. The first chamber houses the pulsed valve (Evan-Levy High-Temperature unmounted valve) with a nozzle of diameter 250 μm at 5 mm from the first skimmer aperture. The third chamber houses the VSI spectrometer, as shown in Figure 1(b). Typical operating pressures in the three chambers with the pulsed valve operating at 5 bar pressure, 1 kHz repetition rate, and 25 μs pulse width are $1 \times 10^{-4}$, $5 \times 10^{-6}$, and $1 \times 10^{-7}$ Torr, respectively. With appropriate delays, the pulsed valve is operated synchronously with the electron gun and pusher pulses.

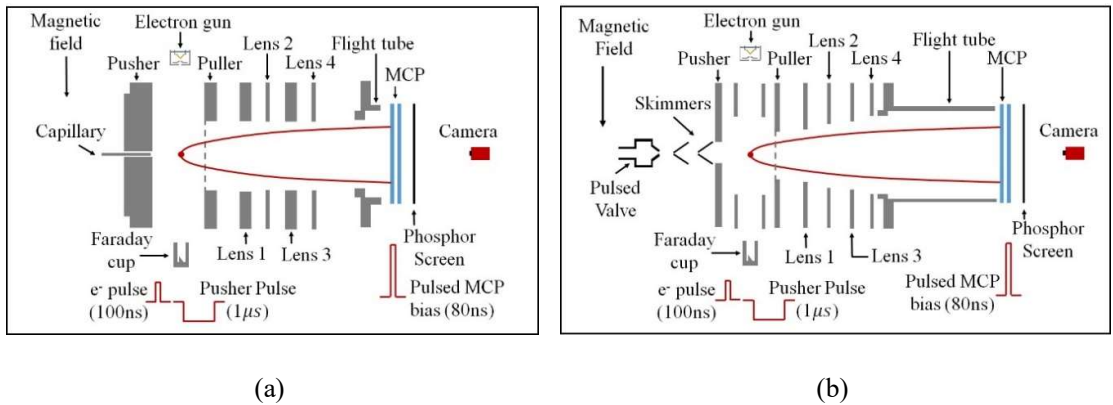

(a)            (b)

FIG. 1. Schematics of the VSI spectrometers used in the experiments with (a) effusive beam (setup-1) and (b) supersonic beam (setup-2) as the target.

**III. SIMULATIONS**

We have carried out the VSI measurements in both setups. We have also performed ion-trajectory simulations using SIMION 8.2 to investigate the spectrometer's performance under various operating conditions for both setups. In this work, we show all images in terms of pixels instead of momentum values to demonstrate the image magnification. However, the kinetic energy and angular distribution are determined from the momentum images, which are obtained by the appropriate transformation of the pixel images.

Electrostatic lenses are best suited to magnify momentum images of the ions generated along the spectrometer axis. The spatial spread in the small interaction volume, the overlapped volume of the molecular and electron beam, can be nullified by the stack of lenses. However, for setup-1, any ion generated in front of the puller electrode aperture (diameter 40 mm) along the path of the electron beam can be extracted to the detector. These ions will experience varying lensing forces based on their position about the spectrometer axis as they would be generated during the electron beam passage through the background gas in the effusive beam set-up (setup-1). To simulate the effect of such ions on the measured momentum images, we have considered the ions source in the simulations as a cylindrical volume with 40 mm length and 1mm diameter along the electron beam path. This cylindrical volume has been kept symmetric about the spectrometer axis. We have used appropriate algorithms for the SIMION platform for applying time-dependent potentials on the electrodes to mimic the delayed extraction potential on the pusher with respect to ion generation. We have also used the initial uncertainty of 100 ns in the ion generation instances to incorporate the effect of the electron pulse width. We note the time of arrival of the ions along with their position on the 2D detector from the simulation for further analysis using a Matlab-based program. Using the ToF of the ions, we obtain several velocity slice images around the centre of the ToF signal with a time window of 80 ns. Among all the images, the image with the largest diameter is chosen as the central slice, as this would correspond to the ions with the maximum velocity in the plane parallel to the detector plane. We have also independently verified the initial momentum distribution of these ions and found it consistent with the Newton sphere's central slice.

In all simulations, we have taken ions of mass 16 amu and initial energy as Gaussian distribution of mean 0.4 eV and FWHM of 0.2 eV, and mean 1.5 eV and FWHM of 0.5 eV. These distributions are similar to those obtained for $O^-$ fragment from DEA to $N_2O$ at 2.3eV and DEA to $O_2$ at 6.5eV, respectively, for a typical incoming electron energy resolution of 1 eV FWHM[18]. The simulations are carried out in two sets: 1) for ions within a sphere of 1mm diameter at the centre of the interaction region, representing the background-free condition (Set-I) and 2) for ions within the above-mentioned cylindrical volume representing the background in the actual experiment (Set-II). In all cases, the initial velocity distribution is taken as isotropic about the origin. The number of ions used for simulations in both sets is consistent with the observed counts from the corresponding regions in the actual experiments.

## IV. RESULTS AND DISCUSSION

We have carried out simulations of ion trajectories for various imaging conditions. We have also carried the actual imaging measurements for the ions generated in DEA measurements under different optimised conditions. In these VSI experiments using the setups described earlier, the slicing is carried out using a pulsed voltage of a fixed width (80 ns). The spread in the ToF signal has some effect on the observed resolution of the image as the fixed-width slicing accesses the central part of the Newton sphere up to different extents depending on the spread in the ToF. Under different voltage conditions, this ToF spread changes affecting the effective momentum resolution.

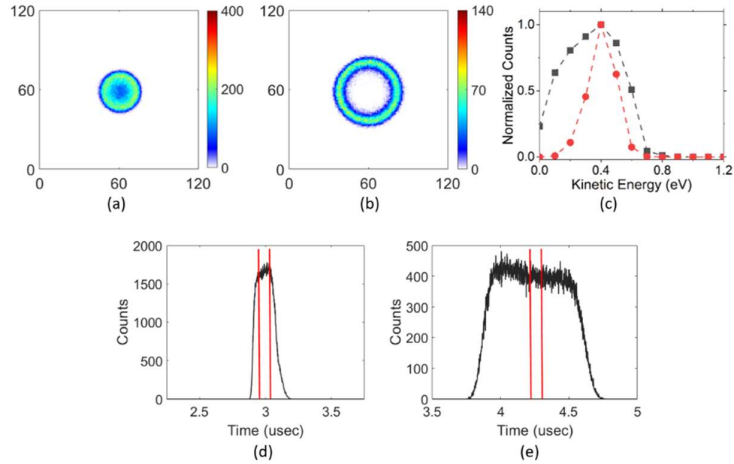

FIG. 2. (a) and (b) are the simulated images for mass 16 amu ions with the initial kinetic energy of 0.4 eV and FWHM of 0.2 eV with the isotropic angular distribution under two different lensing conditions used for the spectrometer in the setup-1 for ions generated from Set-I (please refer to the text). (c) shows the corresponding kinetic energy distribution obtained from these images. Squares (■) for the image (a) and circles (●) for image (b). (d), (e) show the ToF spread of ions, and the red lines show the 80ns slice which corresponds to pixel image (a) and (b) respectively.

Figure 2 shows the simulated slice images for the ions with the initial kinetic energy of 0.4eV from Set-I. Figures 2(a) and 2(b) are obtained for two different lensing conditions for image magnifications. Figure 2(c) shows the kinetic energy distribution obtained from the slice images. The magnified image gives a better kinetic energy distribution, consistent with the initial kinetic energy distribution of about 0.2 eV around 0.4 eV. This shows the effect of spread in the ToF on the imaging using a fixed-width slicing technique.

Below we describe the various schemes we have implemented to minimise the effect of background signal on the slice images.

## A. Minimising the effect of the background using lensing condition

Due to the extended nature of the ion source, a spectrometer's velocity-focusing conditions need to be tweaked by playing with the potentials on various electrodes. We have performed this operation to obtain the best possible image that would have the minimum effect from the background. Figures 3 and 4 show the results of such simulations for the ions of mass 16 amu created with isotropic velocity distribution with the initial kinetic energy of 0.4 eV (FWHM 0.2 eV) and 1.5 eV (FWHM of 0.5 eV), respectively, for three different magnifications.

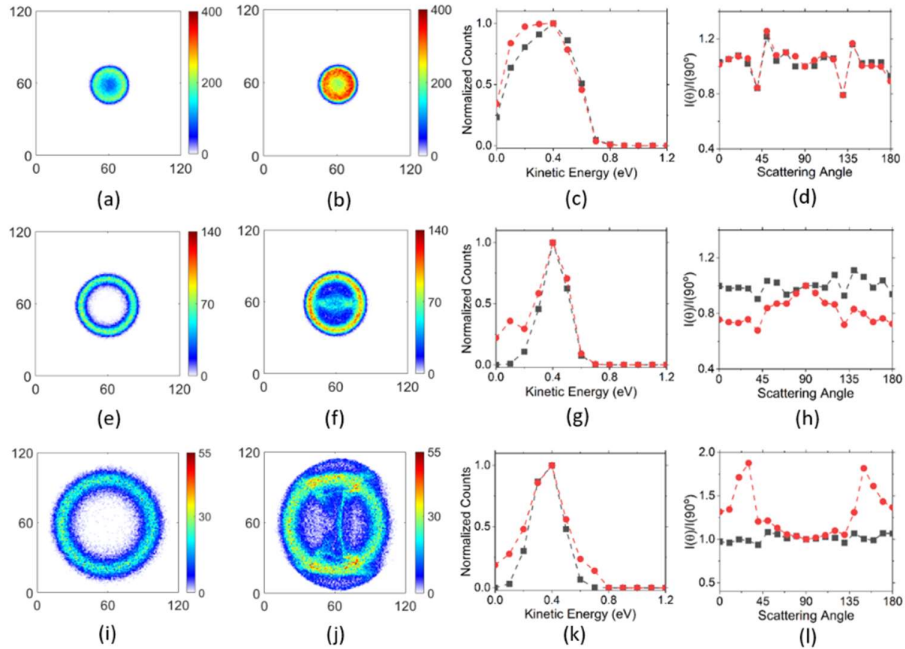

FIG. 3. Simulation result of VSI of mass 16 amu ions with 0.4 eV kinetic energy and 0.2 eV FWHM with the isotropic angular distribution in setup-1. (a), (e) and (i) show the image in the absence of the background under different lensing conditions. (b), (f), and (j) are the same images in the presence of a background. (c), (g), and (k) show the comparison of the kinetic energy distribution in the presence and in the absence of the background. (d), (h), and (l) show the comparison of the angular distribution for the respective cases. In (c), (d), (g), (h), (k), and (l) the squares (■) are for images (a), (e), and (i) and the circles (●) for images (b), (f), and (j). All images are plotted in pixels.

Figures 3 (a), (e), and (i) show the images obtained for the ions of 0.4 eV kinetic energy from the Set-I at three different magnification conditions, whereas Figures 3 (b), (f), and (j) show the images simulated for ions of same energy but with Set-I and Set-II together, mimicking the actual experimental situation. Figures 3 (c), (g) and (k) show the kinetic energy distribution obtained from the corresponding images integrated over all angles. The corresponding angular distributions are shown in Figures 3 (d), (h) and (l). The angular distribution and kinetic energy distribution show the effect of the background signal, which becomes worse for the magnified images. The images themselves show artefacts for higher magnifications arising due to signals from the background gas. The angular distributions and kinetic energy distributions are used to deduce information about the molecular dynamics underneath the dissociation processes. These simulations show the limitations of such imaging methodologies due to background gas present in the apparatus. From these simulations, we infer that for imaging low-energy ions, we have to minimise the background. We also note that these artefacts are imaging condition-dependent. Even if we achieve the same magnification using different lensing conditions, they also generate different artefacts. Here we have shown only one lensing condition per magnification.

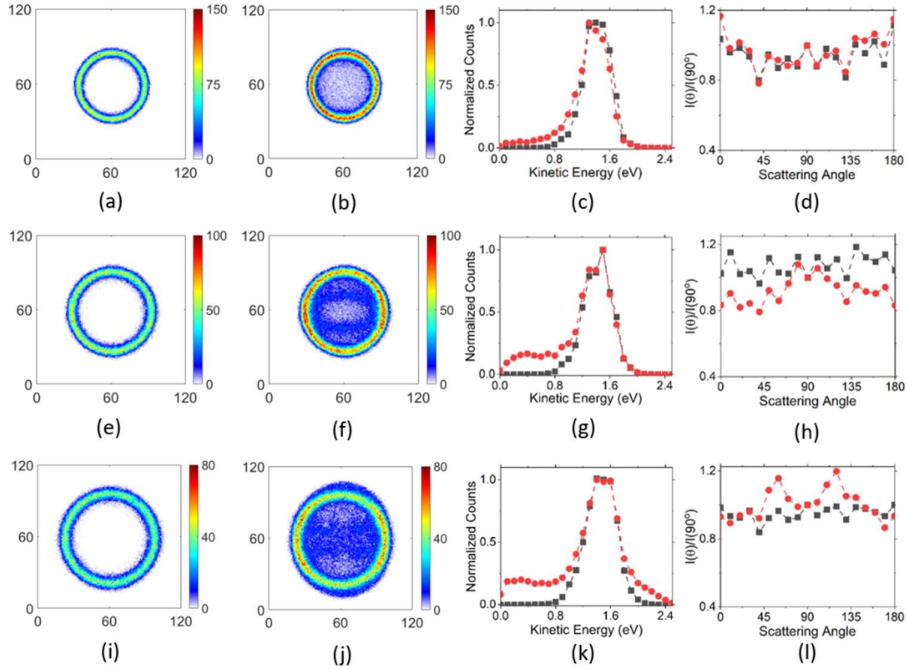

FIG. 4. Simulation result of VSI of mass 16 amu ions with 1.5 eV kinetic energy and 0.5 eV FWHM with the isotropic angular distribution in setup-1. (a), (e) and (i) show the image in the absence of the background under different lensing conditions. (b), (f), and (j) are the same images in the presence of a background. (c), (g), and (k) show the comparison of the kinetic energy distribution in the presence and in the absence of the background. (d), (h), and (l) show the comparison of the angular distribution for the respective cases. In (c), (d), (g), (h), (k), and (l), the squares (■) are for images (a), (e), and (i) and the circles (●) for images (b), (f), and (j). All images are plotted in pixels.

Similar simulations are also carried out for the ions of mass 16 amu created with isotropic velocity distribution with the initial kinetic energy of 1.5 eV and FWHM of 0.5 eV. The results are shown in Figure 4. For ions with higher initial kinetic energies, here, 1.5 eV, we obtain a decent-quality image. However, we should be careful in choosing the imaging condition, as shown in Figure 4, among three images, only condition-1 produces (Figures 4 (a) and (b)) the image, where the image without background matches fairly well with the image in the presence of background in terms of kinetic energy and angular distribution (Figures 4 (c) and (d)).

As can be seen from these simulations, for ions with a given initial energy, depending on the spectrometer geometry, we can obtain the optimised voltage condition where the spatially extended source from the background causes minimum distortion to the slice image. However, this situation worsens for different magnifications. We have also found that the optimised voltage condition for a given magnification that minimises the effect of background is applicable for only a limited initial kinetic energy range (typically up to 2 eV) for a given ion. This is a very difficult solution to implement in practice for the ions generated with a wider range of initial kinetic energy, which is the case for many polyatomic molecules. In such cases, the newly found imaging condition for different kinetic energy ranges would need a fresh calibration.

**B. Subtracting the background contribution**

Another possible way of addressing this issue could be making the measurements of the background contribution and then subtracting it with an appropriate normalisation. We implemented this scheme experimentally in the DEA reaction in $O_2$. We have measured the VSIs for $O^-$ from $O_2$ obtained from DEA at 6.5 eV electron energy[18, 29]. The reaction channel is

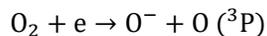
$$O_2 + e \rightarrow O^- + O\ (^3P)$$

Angular distribution of $O^-$ shows four lobes with very low counts at 0º and 180º with respect to electron beam direction[18]. The images obtained for different magnifications are shown in Figure 5. The electron beam direction in all these measurements is from top to bottom. Due to the presence of the transverse magnetic field, the $O^-$ ions trajectories bend in one direction. This shifts the momentum image to one side of the spectrometer axis and introduces distortion[28]. This distortion is particularly starker for higher kinetic energy ions as they tend to travel farther from the spectrometer axis. This is the reason for not having similar intensity on the left and right sides of the image. For the image analysis, we have considered only half of the image obtained close to the centre of the detector, which is distorted the least.

For the first data (Figure 5 (g)), voltages on the electrodes are optimised such that the image can be taken under the best spatial focusing condition. The image shows similar angular distribution as reported by earlier studies[18]. We have changed the electrode voltages to magnify the images ((Figure 5 (h) and (i)). The magnified images show some artefacts, and as we increase the magnification, these artefact patterns change. Based on our simulations, we infer that the artefacts observed in the magnified images are mainly from the background and that the flat angular distribution obtained in Figure 5 (i) is due to the heavy accumulation of artefacts in 90º and 270º directions. Figure 5 (a), which is considered as best space-focusing condition, does not show any artefacts due to the perfect superposition of the background (Figure 5 (b)) contribution with the image obtained from the main beam. In this setup, background contribution is almost 1/3 of the total count coming in the presence of an effusive beam.

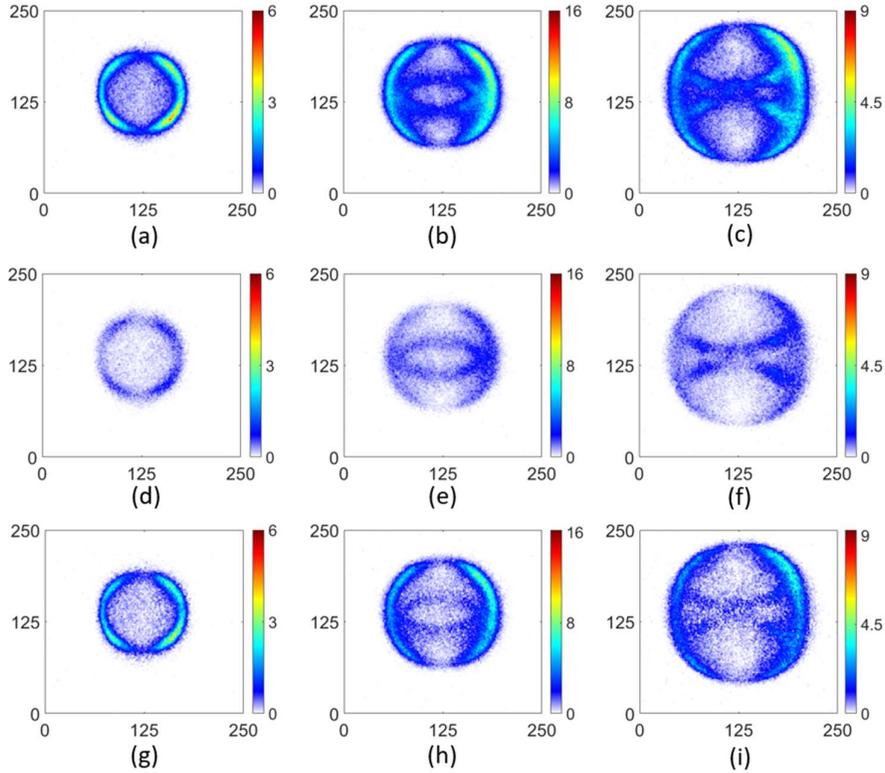

FIG. 5. VSIs obtained for O⁻ from $O_2$ from DEA at 6.5 eV. (a), (b) and (c) the images obtained in the crossed-beam mode under three different lensing conditions. (d), (e) and (f) the images obtained in the static gas mode (background) under the same conditions. (g), (h) and (i) the background-subtracted images obtained by subtracting images (d), (e), and (f) from images (a), (b), and (c), respectively, after appropriate normalisation.

However, even after having very good statistics, these subtracted images are only partially free of background effects. This is mainly because we mimic the background gas contribution in the experiment by recording images for the static gas mode of operation. This condition is fixed by flooding the vacuum chamber with the target gas at the same pressure as that is measured by the ionisation gauge while carrying out the crossed-beam measurements. This is not necessarily an accurate method, as the gauge is mounted far away from the interaction region. Since the interaction region cannot have the same effective pumping speed as that obtained in the part of the vacuum chamber where the gauge is mounted, the contribution from the extended region during the crossed-beam measurement would always be higher. We see this in the subtracted images in Figure 5.

Figure 6 (b) shows the kinetic energy distribution of O⁻ for all three conditions, and the width of the distribution is the same for all of them. This shows that with increasing magnification, we are still in a good spacial focusing condition, but condition 2 and 3 is not good enough to focus the ions generated in the extended interaction region. This is consistent with our discussions in section A. Moreover, the angular distributions of all three subtracted images (Figure 6 (a)) are different. The one with the best-focusing condition shows the angular distribution nearest to the earlier reports. This implies that care needs to be taken while interpreting such data when the background signal is subtracted.

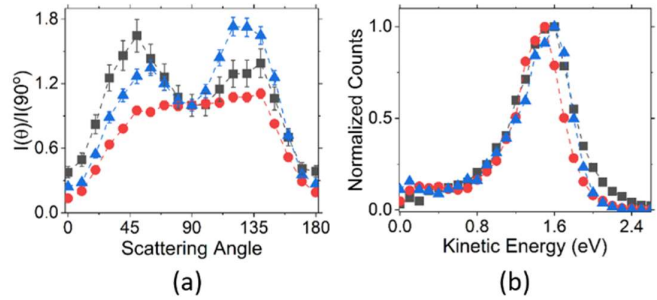

FIG. 6. (a) Angular distribution of $O^-$ from $O_2$ at 6.5 eV for three images obtained in setup-1 after background subtraction given in Figure 5 and (b) corresponding kinetic energy distributions. The squares (■) are for image (g), the circles (●) are for image (h), and the triangles (▲) are for image (i) from Figure 5.

## C. Reducing the background gas contribution.

In both the schemes suggested earlier, we see that the background gas contribution to the image severely limits the performance of the VSI spectrometer for the charged particle collision studies. The third possible solution for such experiments would be to reduce the contribution from the background gas by modifying the target properties. Here, we use the supersonic molecular beam as a target with a well-defined relatively higher density region of the molecular beam, and the corresponding background gas has a very small number density. This we can see from the ion gauge pressure reading of the spectrometer chamber when the molecular beam is on and off. In our experiment, in the effusive beam setup, the background pressure increases from $10^{-8}$ Torr to $10^{-6}$ Torr, whereas in the supersonic molecular beam setup, the pressure increases from $10^{-8}$ Torr to $10^{-7}$ Torr, at least an order of magnitude difference in the two cases, but their overall count rate remains almost identical.

To test the magnification capability and background effect, we have obtained the VSIs of $O^-$ from DEA to $N_2O$ and $O_2$ in both setups. The reaction channel for DEA to $N_2O$ at 2.3 eV is

$$N_2O + e \rightarrow O^- + N_2(X^1\Sigma_g^+)$$

$O^-$ has a mean kinetic energy of 0.4 eV[30] with a spread of ±0.2 eV, it shows a nice ring in the pixel image. The images obtained from both setups with two different magnifications are shown in Figure 7. For setup-1, as we can see from Figure 7 (a), we have obtained a nice ring consistent with the reported data. However, on magnifying it, the image loses all of these features (Figure 7 (b)). This is consistent with the simulation results presented earlier. We understand this as due to the contribution from the background gas. We have imaged the same ions in setup-2, where the target gas is a supersonic molecular beam which has the background gas an order of magnitude lower in pressure as compared to the effusive beam of setup-1. We have taken images for two different magnification conditions, the same as for setup-1 (Figure 7 (c) and (d)). Unlike setup-1, in this case, both images show similar distributions. These findings are also consistent with our simulation results confirming that by suppressing background, we can easily magnify images of low initial kinetic energy.

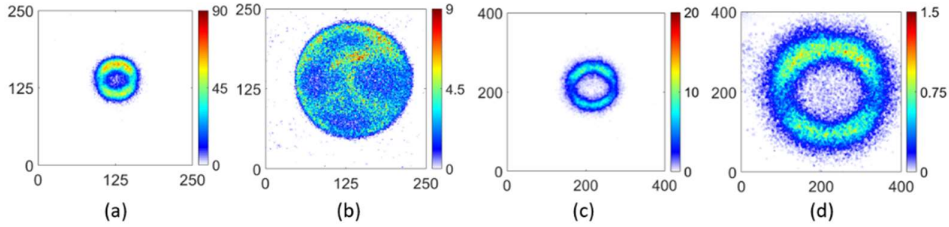

FIG. 7. Experimentally obtained VSIs of $O^-$ from $N_2O$ at 2.3eV electron energy. (a), and (b) are taken in setup-1 and (c), and (d) are taken in setup-2 under different magnification conditions, respectively.

We further investigate the magnification capability and the effect of background by imaging the $O^-$ ions from DEA to $O_2$. In this scheme, first, the electrode voltages are optimised to get the best spatially focused image (Figure 8 (a)). Then the image is magnified by changing electrode voltages (Figure 8 (b) and (c)). All three images have very little contribution from the background. Here, the transverse magnetic field distorts the right side of the image, and this setup has a longer flight tube, $O^-$ ions fly over a longer distance and hence for a longer time in the magnetic field. This causes more prominent distortion of the image compared to setup-1, which has a shorter flight tube.

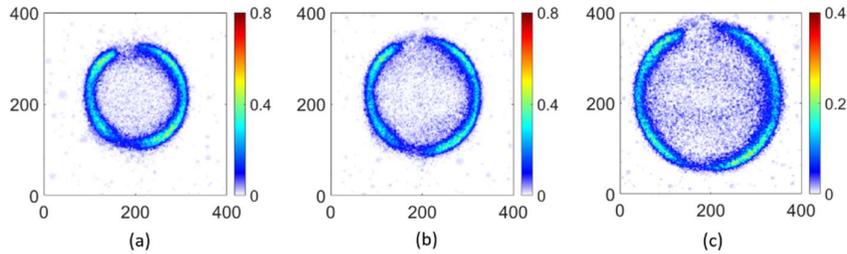

FIG. 8. (a), (b), and (c) shows the images of $O^-$ from $O_2$ from DEA at 6.5 eV in the crossed-beam mode under three different lensing conditions.

The angular distribution shows a similar pattern for all three conditions (Figure 9 (a)). The kinetic energy distribution also shows a similar spread which shows all of the images are in the best special focusing condition. This also shows that we can magnify images without compromising the momentum resolution. We have also found that the detector size limits the extent of magnification achieved for the spectrometer geometry used in these experiments. Kinetic energy distribution for setup-2 images is narrower than for setup-1 images. This is due to the low thermal energy spread in molecules. Supersonic beams produce an overall superior image compared to effusive beams under all magnifications.

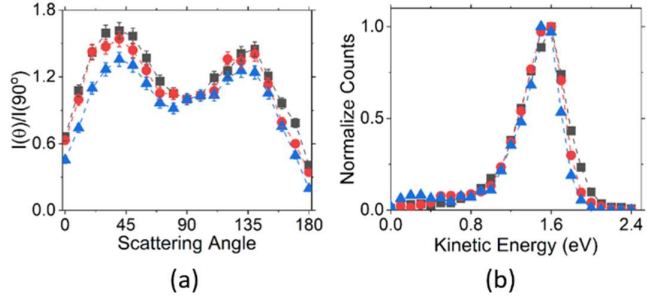

FIG. 9. (a) Angular distribution of O⁻ from $O_2$ at 6.5 eV for all three images obtained in setup-2 in the crossed-beam mode shown in Figure 8, and (b) corresponding kinetic energy distributions. The squares (■) are for image (a), the circles (●) are for image (b), and the triangles (▲) are for image (c) from Figure 8.

## V. CONCLUSION

In this work, we have shown how the presence of background gas creates artefacts in the velocity slice imaging, particularly in the charged particle interaction studies where the projectile beam is not focused. These artefacts change with the magnification of the momentum imaging. Hence magnifying an image in a crossed-beam setup, where the effusive beam is used as the target source, is a challenging task. If not done carefully, these artefacts can lead to an erroneous interpretation of data. This problem can be solved by minimising the background gas. One way to achieve this is by using a supersonic jet as the target beam, which shows cleaner images under all magnification and a narrower kinetic energy spread. However, generating a supersonic molecular beam with sufficient number density for molecules with a very low vapour pressure is a challenge in such schemes. The situation would be even more difficult for studying processes like DEA, which have relatively lower cross-sections. In such a scenario, an effusive molecular beam is easier to work with. In that case, appropriate focusing condition needs to be worked on. For a moderate to high kinetic energy range, we can operate the spectrometer in a medium magnification range where the static gas contribution has minimum effect on the crossed-beam images overlap.

**ACKNOWLEDGMENTS**

S.D. and V.S.P. acknowledge the financial support from the Department of Atomic Energy, India, under Project Identification No. RTI4002. S.D. wishes to thank S. Swain for teaching the operations of the experimental setup and analysis processes. S.D. also thanks S. Tare and Y. Upalekar for their technical support.

# REFERENCES

1. D. W. Chandler, and P. L. Houston, "Two-dimensional imaging of state-selected photodissociation products detected by multiphoton ionization," J. Chem. Phys. **87**(2), 1445–1447 (1987). https://doi.org/10.1063/1.453276
2. A. T. J. B. Eppink and D. H. Parker, "Velocity map imaging of ions and electrons using electrostatic lenses: Application in photoelectron and photofragment ion imaging of molecular oxygen," Rev. Sci. Instrum. **68**(9), 3477–3484 (1997). https://doi.org/10.1063/1.1148310
3. H. L. Offerhaus, C. Nicole, F. Lépine, C. Bordas, F. Rosca-Pruna, and M. J. J. Vrakking, "A magnifying lens for velocity map imaging of electrons and ions," Rev. Sci. Instrum. **72**(8), 3245–3248 (2001). https://doi.org/10.1063/1.1386909
4. C. Bordas, F. Lépine, C. Nicole, and M. J. J. Vrakking, "Photoionization Microscopy," Phys. Scr. **68**, (2004). https://doi.org/10.1238/Physica.Topical.110a00068
5. S. Cohen, M. M. Harb, A. Ollagnier, F. Robicheaux, M. J. J. Vrakking, T. Barillot, F. Lépine, and C. Bordas, "Wave Function Microscopy of Quasibound Atomic States," Phys. Rev. Lett. **110**, 183001 (2013), https://doi.org/10.1103/PhysRevLett.110.183001
6. A. S. Stodolna, A. Rouzée, F. Lépine, S. Cohen, F. Robicheaux, A. Gijsbertsen, J. H. Jungmann, C. Bordas, and M. J. J. Vrakking, "Wave Function Microscopy of Quasibound Atomic States," Phys. Rev. Lett. **110**, 213001 (2013), https://doi.org/10.1103/PhysRevLett.110.183001
7. N. H. Abel, Journal fur die reine und agewandte Mathematik, 1, pp. 153-157 (1826).
8. A. J. R. Heck, and D. W. Chandler, "Imaging techniques for the study of chemical reaction dynamics," Annu. Rev. Phys. Chem. **46**, 335 (1995). https://doi.org/10.1146/annurev.pc.46.100195.002003
9. C. Bordas, F. Paulig, H. Helm, and D. L. Huestis, "Photoelectron imaging spectrometry: Principle and inversion method," Rev. Sci. Inst. **67**, 2257 (1996). https://doi.org/10.1063/1.1147044
10. M. J. J. Vrakking, "An iterative procedure for the inversion of two-dimensional ion/photoelectron imaging experiments," Rev. Sci. Instrum. **72**, 4084 (2001). https://doi.org/10.1063/1.1406923
11. V. Dribinski, A. Ossadtchi, V. Mandelshtam, and H. Reisler, "Reconstruction of Abel-transformable images: The Gaussian basis-set expansion Abel transform method," Rev. Sci. Instrum., **73**, 2634–2642 (2002). https://doi.org/10.1063/1.1482156
12. G. A. Garcia, and L. Nahon, I. Powis, "Two-dimensional charged particle image inversion using a polar basis function expansion," Rev. Sci. Instrum. 75, 4989-4996 (2004). https://doi.org/10.1063/1.1807578
13. C. Sparling, A. Ruget, J. Leach, and D. Townsend, "Arbitrary image reinflation: A deep learning technique for recovering 3D photoproduct distributions from a single 2D projection," Rev. Sci. Instrum. **93**, 023303 (2022). https://doi.org/10.1063/5.0082744
14. C. Sparling, A. Ruget, N. Kotsina, J. Leach, and D. Townsend, "Artificial Neural Networks for Noise Removal in Data-Sparse Charged Particle Imaging Experiments," Chem. Phys. Chem. **22**, 76-82 (2021). https://doi.org/10.1002/cphc.202000808
15. C. R. Gebhardt, T. P. Rakitzis, P. C. Samartzis, V. Ladopoulos, and Kitsopoulos, "Slice imaging: A new approach to ion imaging and velocity mapping," Rev. Sci. Instrum. **72**, 3848-53 (2001). https://doi.org/10.1063/1.1403010

16. D. Townsend, M. P. Minitti, and A. G. Suits, "Direct current slice imaging," Rev. Sci. Instrum. **74**, 2530–2539 (2003). https://doi.org/10.1063/1.1544053
17. J. J. Lin, J. Zhou, W. Shiu, and K. Liu, "Application of time-sliced ion velocity imaging to crossed molecular beam experiments," Rev. Sci. Instrum. **74**, 2495–2500 (2003). https://doi.org/10.1063/1.1561604
18. D. Nandi, V. S. Prabhudesai, E. Krishnakumar, and A. Chatterjee, "Velocity slice imaging for dissociative electron attachment," Rev. Sci. Instrum. **76**, 053107 (2005). https://doi.org/10.1063/1.1899404
19. B. Wu, L. Xia, H. K. Li, X. J. Zeng, and S. X. Tian, "Positive/negative ion velocity mapping apparatus for electron-molecule reactions," Rev. Sci. Instrum. **83**, 013108 (2012). https://doi.org/10.1063/1.3678328
20. A. Moradmand, J. B. Williams, A. L. Landers, and M. Fogle, "Momentum-imaging apparatus for the study of dissociative electron attachment dynamics," Rev. Sci. Instrum. **84**, 033104 (2013). https://doi.org/10.1063/1.4794093
21. E Szymariska, V. S. Prabhudesai, N. J. Mason, and E. Krishnakumar, "Dissociative electron attachment to acetaldehyde, $CH_3CHO$. A laboratory study using the velocity map imaging technique," Phys. Chem. Chem. Phys. **15**, 998-1005 (2013). https://doi.org/10.1039/C2CP42966G
22. H. Adaniya, B. Rudek, T. Osipov, D. J. Haxton, T. Weber, T. N. Rescigno, C. W. McCurdy, and A. Belkacem, "Imaging the Molecular Dynamics of Dissociative Electron Attachment to Water," Phys. Rev. Lett. **103**, 233201(2009). https://doi.org/10.1103/PhysRevLett.103.233201
23. P. Nag, M. Polášek, and J. Fedor, "Dissociative electron attachment in NCCN: Absolute cross sections and velocity-map imaging," Phys. Rev. A **99**, 052705 (2019). https://doi.org/10.1103/PhysRevA.99.052705
24. G. Li, H. J. Hwang, and H. C. Jung, " High resolution kinetic energy by long time-delayed core-sampling photofragment translational spectroscopy," Rev. Sci. Instrum. **76**, 023105 (2005). https://doi.org/10.1063/1.1844412
25. V. Plomp, Z. Gao, and M. Sebastiaan, "A velocity map imaging apparatus optimised for high-resolution crossed molecular beam experiments," Mol. Phys. **119** e1814437 (2021). https://doi.org/10.1080/00268976.2020.1814437
26. J. O. F. Thompson, C. Amarasinghe, C. D. Foley, and A. G. Suits. "Finite slice analysis (FINA)-A general reconstruction method for velocity mapped and time-sliced ion imaging," J. Chem. Phys. **147**, 013913 (2017). https://doi.org/10.1063/1.4979305
27. A. Moradmand, J. B. Williams, A. L. Landers, and M. Fogle, "Momentum-imaging apparatus for the study of dissociative electron attachment dynamics," Rev. Sci. Instrum. **84**, 033104 (2013). https://doi.org/10.1063/1.4794093
28. S. Swain, E. Krishnakumar, and V. S. Prabhudesai, "Structure and dynamics of the negative-ion resonance in $H_2$, $D_2$, and HD at 10 eV," Phys. Rev. A **103**, 062804 (2021). https://doi.org/10.1103/PhysRevA.103.062804

29. P. J. Chantry, and G. J. Schulz, "Kinetic-Energy Distribution of Negative Ions Formed by Dissociative Attachment and the Measurement of the Electron Affinity of Oxygen," Phys. Rev. **156**, 134 (1967). https://doi.org/10.1103/PhysRev.156.134
30. D. Nandi, V. S. Prabhudesai, and E. Krishnakumar, "Dissociative electron attachment to $N_2O$ using velocity slice imaging," Phys. Chem. Chem. Phys. **16**, 3955-3963 (2014). https://doi.org/10.1039/C3CP53696C